

\input amstex
\input vanilla.sty
\nopagenumbers
\baselineskip 16pt
\magnification=1200
\def\D{\Cal{D}}
\def\np{\not{\text{P}}}
\hsize=16 true cm
\vsize=23 true cm

\TagsOnRight
\def\udt#1{$\underline{\smash{\hbox{#1}}}$}
\def\wh{\widehat}
\def\wt{\widetilde}

\font\catorce=cmr10 scaled 1440
\def\ov{\overline}

\def\noi{\noindent}
$\ $
\vskip 4cm
\centerline{\bf COVARIANT QUANTIZATION OF}
\centerline{\bf GREEN SCHWARZ SUPERSTRING}
\vskip 1cm

\centerline{A. Restuccia and J. Stephany}

\vskip 5mm

\centerline{\it Universidad Sim\'on Bol\'{\i}var, Dept. de F\'{\i}sica,
Caracas}

\vskip 1cm

\centerline{\bf Abstract}

{\narrower\flushpar{We describe a canonical covariant approach to the
quantization of the Green-Schwarz superstring. The  approach is first
applied to the canonical covariant quantization of the Brink and Schwarz
superparticle. The Kallosh action is obtained in this case, with the correct
BRST cohomology.}\par}

\newpage

\noi
{\bf 1. Introduction}

\vskip 3mm

The construction of the second quantized, Lorentz covariant Superstring
action is one of the main problems to be solved in String Theory. The Lorentz
covariant, BRST quantization of the Green-Schwarz Superstring
(GSSS) [1] is the first step we have to take towards that goal.

A non-perturbative, second quantized action for superstrings was formulated
by Green-Schwarz [2] in the Light Cone Gauge (LCG). The action is cubic in
terms of the functions of the transverse modes of the superstrings with a non
trivial insertion factor. It was shown in [3] that the action requires in
addition complicated contact terms in order to allow the closure of the
Super-Poincare algebra. Recently a new approach to Superstring in the LCG has
been proposed [4]. It is obtained by patching together the local
Green-Schwarz LCG action, using the unique complex analytic structure of
Teichm\"uller space. It may solve completely the problem of contact terms at
the second quantized level for closed type IIB and Heterotic
strings. However, in order to understand the geometric principles of the
Superstring  Theory a Lorentz covariant formulation is required. As
usually we are interested in a covariant, canonical, BRST quantization of
the  Green-Schwarz Superstring [1], in order to have simultaneously a
manifest covariant formulation and unitary theory. Given a gauge
theory whose  evolution is determined by a Hamiltonian $H$ on a phase space
$\Cal{M}$ of  local coordinates $(q,p)$ with first class, irreducible,
regular constraints  $\varphi_i$, then the construction of the effective
action follows by  considering the BRST charge $Q$,
$$\align
\{Q,Q\}_{Poisson} &=0,\\
Q|_{\mu =0} &= c^i\varphi_i,\tag 1.1
\endalign
$$
where $(c^i,\mu_i)$ are the ghost fields and its conjugate momenta. The BRST
invariant Hamiltonian $\wh{H}$ is obtained from
$$\align
\{\wh{H},Q\}&=0,\\
\wh{H}|_{\mu =0} &= H.\tag 1.2
\endalign
$$
The effective action is given by
$$
S_{eff}=<p\dot{q}+\mu \dot{c}-\wh{H}+\wh{\delta}(\lambda^i\mu_i)+\wh{\delta}
(\ov{c}_i\chi^i)>,\tag 1.3
$$
where $\wh{\delta}$ denotes BRST transformed [5], and $\chi^i$ are the gauge
fixing conditions.

The functional integral
$$
I(\chi )=\int \Cal{D}_\mu exp\frac{i}{k}S_{eff},\tag 1.4
$$
whith $\Cal{D}_\mu$ the Liunville measure, is then independent of
$\chi$, that is,
$$
\frac{\delta I}{\delta\chi}=0 \tag 1.5
$$
within the admissible set of gauge fixing conditions, provided initial and
final (in time) conditions on the ghost fields are satisfied [5].

Property (1.5) allows to shown the equivalence between the quantization of the
gauge theory in a physical gauge, like the LCG where only physical modes are
present and where physical unitary is directly proven, and the manifest
Lorentz covariant quantization obtained in a covariant gauge, for example
$\lambda_0=1$ $\lambda_1=0$ in the Bosonic String Theory. $\lambda_0$ is
the  Lagrange multiplier (L.m.) associated to the Hamiltonian constraint
while  $\lambda_1$ is the L.m. associated to the generator for
$\sigma$-reparametrizations.

The construction of $S_{eff}$ for the GSSS and property (1.5) for a
non-trivial topology are the ingredients for a proof of the equivalence
between the  perturbative multiloop amplitudes associated of the theory in
the LCG and in a Lorentz Covariant gauge. The corresponding  equivalence
for the bosonic string theory following this line was done in [6]. The
construction of the covariant, BRST effective action for the GSSS in thus
relevant in several aspects. The main difficulty in this construction has
been the covariant gauge fixing  of the local $K$ supersymmetry. The first
class constraints  associated with the gauge symmetries appear mixed with
second class ones, and  so far no local, Lorentz covariant, and finite
reducible approach to disentangle them has been found.

The same difficulties, but at less sophisticated level, are present in the
covariant quantization of the Brink-Schwarz superparticle (BSSP) which
correspond to  the zero mode of the GSSS. For this reason all approaches to
the covariant quantization of superstrings  have been first tested with the
BSSP.

The approach we are going to follow in order to obtain the covariant, BRST
invariant, effective action for the GSSS was developed in Ref. [7]. It is
based on the idea [8] of extending a theory with second class constraints to
another one with only first  class constraints  in such a way that the
functional integral of the  latest reduces by partial gauge fixing, within
the admissible set, to the  original one with the correct Fradkin-Senjanovic
measure. It is an off-shell  approach allowing the systematic construction
of the off-shell nilpotent BRST  charge and of the BRST invariant effective
action. In the extended phase  space where all constraints are first class
the operatorial quantization  construction of Batalin and Fradkin [5] may be
directly implemented. However for the GSSS as well as for the BSSP due  to
the infinite reducibility structure of the first class constraints,
generating functions should be introduced in order to handle the infinite
set  of auxiliary fields. An analogous situation occurs in the
construction of the covariant unconstrained formulation of Super
Yang-Mills and Supergravity [9].

In section 2 we present the GSSS action and perform the canonical analysis,
which was first done in [10]. In section 3 discuss the structure of the
constraints in relation with the previous approaches to deal with the
problem,  which involved the introduction of irregular constraints [11].In
section 4 we present our method applied to the covariant quantization of the
superparticle. We  obtain in a canonical way the action proposed by Kallosh
action [12]. Finally in section 5 we show the construction for the GSSS and
discuss  related problems.

\vskip 5mm
\noi
{\bf 2. The Green-Schwarz covariant Superstring}

\vskip 3mm

The Green-Schwarz covariant action for Superstring Theory in 10 dim may be
written as
$$
S={1\over{\pi}}\int d\sigma d\tau (L_1+L_2+\lambda (det\  \ e+1)),\tag 2.1
$$
where
$$\align
L_1=&-{1\over{2}}e^{\alpha\beta}\pi_\alpha{}^\mu \pi_{\mu\beta}\\
L_2=&-\epsilon^{\alpha\beta}\partial_\alpha X^\mu
\{K^1_{\mu\beta}-K^2_{\mu\beta}\}-\epsilon^{\alpha\beta}K^{1\mu}_\alpha
K^2_{\mu\beta}\\
\pi^\mu_\alpha =&\partial_\alpha X^\mu -\sum_Ai\bar{\theta}^A\gamma^\mu
\partial_\alpha \theta^A\\
K^{A}{}^\mu_\alpha =&i\bar{\theta}^A\gamma^\mu \partial_\alpha \theta^A
\tag 2.2
\endalign
$$
$\theta^A$, $A=1,2$, are Majorana Weyl spinors in ten dimensions, $\alpha$,
$\beta$ denote the world sheet indices.

(2.1) is the functional action in terms of the independent variables $X^\mu$,
$\theta^A$, and $e^{\alpha\beta}$.

The action has manifest Poincare invariance in 10-dim and local
re\-pa\-ra\-me\-tri\-za\-tion invariance on the world sheet. Additionally it
has global and local Supersymmetries that we discuss later on using a
canonical approach.Here we are going to discuss explicitly the closed
Superstring Theory and then  comment on the results for Heterotic String
Theory in 10-dim.

We impose periodic boundary conditions on $X^\mu$, $\theta^A$,
$e^{\alpha\beta}$ at $\sigma =0$ and $\sigma =\pi$ and denote $p_\mu$,
$p_{\alpha\beta}$ and $\xi^A$ the canonical conjugate  momenta associated
to $X^\mu$, $e^{\alpha\beta}$ and $\theta^A$ respectively.

The primary constraints are
$$\align
p_{\alpha\beta}=&0,\tag 2.3a\\
F^A\equiv &\xi^A+i\bar{\theta}^A\gamma^\mu \{p_\mu +(-1)^A{1\over{\pi}}
(\partial_\sigma X_\mu -K_\mu{}^A{}_\sigma )\}=0\tag 2.3b
\endalign
$$

By solving det $e+1=0$ and using the conservation of (2.3) one may get rid of
$e^{\alpha\beta}$ and $p_{\alpha\beta}$ as canonical variables. The
Hamiltonian is the given by
$$
H=\int d\sigma \{-{1\over{e_{\sigma\sigma}}}\phi_1+{e_{\tau\sigma}
\over{e_{\sigma\sigma}}}\phi_2+\wt{F}^1\Lambda_1+\wt{F}^2\Lambda_2\} ,
\tag 2.4
$$
where
$$\align
\phi_1\equiv &{1\over{2\pi}}(\pi^2\wt{p}^2+\pi^2_\sigma -2\pi F^1
\partial_\sigma \omega^1+2\pi F^2\partial_\sigma \theta^2)\\
=&{1\over{2}}(\pi p^2+x^2{}')-\xi^1\theta^1{}'+\xi^2\theta^2{}',\tag
2.5a\\  \phi_2\equiv &\wt{p}\pi_\sigma +F^1\partial_\sigma \theta^1+
F^2\partial_\sigma \theta^2=\\
=&px'+\xi^1\theta^1{}'+\xi^2\theta^2{}',\tag 2.5b\\
\wt{F}^A\equiv &F^A\Gamma_A .\tag 2.5c
\endalign
$$

We have denoted
$$\align
\wt{p}_\mu\equiv &p_\mu +{1\over{\pi}}(K^1_{\mu\sigma}-K^2_{\mu\sigma}),\\
\Gamma_A\equiv &\gamma^\mu (\pi\wt{p}_\mu+(-1)^A\pi_{\sigma\mu}),
\endalign
$$
and repetition of $A$ indices does not denote summation.

Eq.(2.4) is subject to the second class fermionic constraints coming from
(2.3b) $$\align
G^1\equiv &F^1\Gamma_2=0,\\
G^2\equiv &F^2\Gamma_1=0.\tag 2.6
\endalign
$$

The Lagrange multipliers $\Lambda_A$ are restricted to the subspace
$$\align
\Gamma_2\Lambda_1=&0\\
\Gamma_1\Lambda_2=&0.\tag 2.7
\endalign
$$

The constraints (2.3b) have been decoupled into a first class part given by
$\wt{F}^A=0$ and a second class part given by $G^A=0$. The Lagrange
multipliers ${1\over{e_{\sigma\sigma}}}$,
${e_{\tau\sigma}\over{e_{\sigma\sigma}}}$ and $\Lambda_A$ cannot be determined
by Dirac approach. They are thus associated to first class constraints.

Eq.(2.6) were first obtained by Hori and Kamimura using the projectors
$$\align
P_1=&{\Gamma_1\Gamma_2\over{2(\wt{p}^2-\pi^2_\sigma )}},\\
P_2=&{\Gamma_2\Gamma_1\over{2(\wt{p}^2-\pi^2_\sigma )}}.
\endalign
$$

We notice however that $P_1$ and $P_2$ are projectors only over the
submanifold  of constraints. That is not the situation on an effective
action where a  covariant gauge fixing condition, through restrictions on
the Lagrange  multipliers, has been imposed. After all we are just
interested in that case. We thus conclude that the approach of Hori and
Kamimura does not solve the  problem of covariant decoupling of the first
and second class constraints in  Superstring Theory.

Expressions(2.5a,b) are the generators of local reparemetrization on the
world sheet, they  satisfy the same algebra as the Virasoro constraints of
the Nambu-Goto String. $X^\mu$ and $\theta^A$ transform as scalars under
reparametrizations. This  transformation law for $\theta^A$ is essential in
the Green-Schwarz  construction of the interacting theory. Since $\theta^A$
behave as scalars  under reparametrization there is \udt{no summation} on
spin structures  involved in the evaluation of Multiloop Amplitudes.

$\wt{F}^A$ generates local supersymmetric transformation. $X^\mu$ and
$\theta^A$ transform as
$$\align
\delta X^\mu (\sigma )=&\{X^\mu (\sigma ),\sum_B\int d\wt{\sigma}\wt{F}^B
(\sigma )\xi^B(\wt{\sigma})\}=\\
=&\sum_Bi\bar{\theta}^B(\sigma )\gamma^\mu \delta\theta^B(\sigma )\tag 2.8a\\
\delta\theta^A(\sigma )=&\{\theta^A(\sigma ),\sum_B\int d\wt{\sigma}\wt{F}^B
(\wt{\sigma})\xi^B(\wt{\sigma})\}=\\
=&\Gamma_A(\sigma )\xi^A(\sigma ).\tag 2.8b
\endalign
$$

\vskip 5mm
\noi
{\bf 3. Regular and Irregular Constraints.}

\vskip 3mm

In order to circumvent the problem presented by mixing, in a covariant
treatment, of the first and second class constraints other actions for
description of the superparticles and GSSS, following original ideas of
Siegel, have been proposed [13]. They allow a formulation in terms of first
class constraints only. The formulations however are given in terms of
irregular constraints. The presence of irregular constraints [11] does not
allow a straightforward application of the Batalin-Fradkin or
Batalin-Vilkovisky approach. As a consequence the BRST invariant effective
action for these models has not been constructed.

The distinction between regular and irregular constraints is very revelant
since all the quantization procedures break-down in the presence of irregular
constraints.

Let $f$ be a $C^\eta$ map between Banach manifolds
$$
f:M\to N
$$
$n\in N$ is a regular value of $f$ if for each $m(\in f^{-1}(n)$ the map
between tangent spaces at $m$, $T_mf$ is surjective. If $T_mf$ is not
surjective, $m\in M$ is a critical point and $n=f(m)$ is a critical value of
$f$.

The constraints on the evolution of dynamical system may be defined from a
map
$$
\phi :M\to N\tag 3.1
$$
between Banach manifolds, as the set of restrictions
$$
\phi (m)=0\ \ \ \ \ ,m\in M\ \ ,0\in N.
$$
When $0\in N$ is a critical value of $\phi$ we say that $\phi =0$ an irregular
constraints. That is, there exist at least one critical point $m\in M$.

If $\phi$ is a smooth submersion then the constraint $\phi =0$ is a regular
constraint.

The regularity of the constraints is a requirement for the application of the
Lagrange Multiplier theorem. Let $\phi$ be a smooth submersion (3.1), denote
$$
L=\phi^{-1}(0)
$$
and $f$ a functional on $M$
$$
f:M\to N.
$$
Then, $m\in M$ is a critical point of $f|_L$ if and only if there exists
$\lambda \in N^*$, the dual to $N$, such that $m$ is a critical point of
$f-\lambda \cdot \phi$.It is not difficult to construct examples with
irregular constraints where a  critical point of $f|_L$ is not a critical
point of $f-\lambda \cdot \phi$.

In the presence of irregular constraints the
Batalin-Fradkin-Vilkovisky [5] theorem is not valid. In fact, since in
general  the Lagrange Multiplier theorem does not apply in the presence of
irregular  constraints one may suspect that the BFV approach which
explicity introduces  the constraints into the effective action through
Lagrange multipliers may have some  obstruction. The problem arises when
writting the Fradkin-Senjanovic measure  on the functional integral. This
is so, since the $\delta$ Dirac delta is ill  defined on an irregular
constraint.In fact,since
$$
\delta (\phi (x))=\Sigma \frac{\delta^{-1}(\phi (0))}
{det\frac{\delta\phi}{\delta x}|_{\phi^{-1}(0)}},
$$
if $\phi$ is irregular the determinant in the denominator is
zero. Consequently one is not allowed to impose a canonical gauge in the
effective  action associated to a system with irregular constraints, since
integration  on the ghost and antighost fields one ends up with a
Fradkin-Senjanovic like  measure
$$
\Cal{D}_\mu \ \ \ \delta (\phi_{irregular})\ \ \ \delta (\chi)\ \ \
det\{\chi ,\phi \}.
$$

It is still possible, however, to consider covariant gauges in the effective
action. In fact a covariant gauge is implemented by imposing restrictions on
the Lagrange multipliers. One is not working then on the submanifold of
constraints and there is no inconsistency in the procedure. Nevertheless, the
equivalence between a manifest covariant quantization and the quantization on
a physical gauge, where physical unitary is manifest, breaks down.

Finally we would like to remark that the Dirac algorithm to obtain the
constraints on a canonical formulation of a gauge theory has to be
supplemented with the following procedure in order to detect the complete
set of constraints. At each step of the conservation procedure, from a set
of constraints $\phi$ one considers the conservation condition
$$
\{H,\phi \}=0\tag 3.2
$$

\noi
a) If $\phi$ is regular, the algorithm stops when $\{H,\phi \}$ is weakly
zero or when only determines Lagrange multipliers, which must then be
associated to second class constraints.

\noi
b) If $\phi$ is irregular, even when the above conditions are satisfied the
algorithm does not stop, one has also to evaluate
$$
\{H,\{H,\phi \}\}=0,\tag 3.3
$$
if $\{H,\phi \}$ is regular then a) applies if it is irregular one has to
continue the conservation procedure.

The above remark is based in the following point. If $\phi =0$ is a regular
constraint
$$
\{H,\phi \}\approx 0
$$
implies
$$
\{\undersetbrace{n}\to{H,\cdots \{H},\phi \}\cdots \}=0
$$
for all $n$.

This remark can be proven as follows; if $F$ is weakly zero $F\approx 0$
and $\phi$ are regular then
$$
F=\Sigma a\phi
$$
 This property is not valid in general when $\phi$
are irregular constraints.

\vskip 5mm

\noi
{\bf 4. Canonical Covariant Quantization of the Brink-Schwarz Superparticle.}

\vskip 3mm

The first order action for the ten dimensional BS superparticle is
$$
S=<P_\mu \partial_\tau \chi^\mu +\ov{\xi} \np \partial_\tau \xi+eP^2>\tag 4.1
$$
where $e$ is a Lagrange multiplier associated to the constraint
$$
P^2=0\ \ \ \ \ .\tag 4.2
$$

Let $\eta$ be the momenta canonically conjugate to $\xi$. Since the action
(4.1) is first order in $\partial_\tau \xi$ its dynamics is restricted by,
$$
\phi =\eta -\np \xi =0 \ \ \ \ \ .\tag 4.3
$$

The canonical Hamiltonian action of the system is
$$
S=<P_\mu \partial_\tau \chi^\mu +\ov{\eta} \partial_\tau \xi +eP^2+\ov{\psi}
(\eta -\np \xi )>\tag 4.4
$$
where $\ov{\psi}$ are Lagrange multipliers associated to the constraints (4.3).

Constraints (4.3) are a combination of first and second class ones. This is
best observed computing the Poisson algebra of the constraints which yields
$$
\{\phi ,\phi \}=2\np \ \ \ \ \ .\tag 4.5
$$
Over the manifold defined by (4.2), $\np$ is non-invertible and in fact
conservation of (4.3) fixes only half of the multipliers $\ov{\psi}$. The
constraints associated to the the other half are first class.

One can covariantly project from (4.3) the first class constraints by
application of $\np$
$$
\varphi \equiv \np \eta = 0 \ \ \ \ \ .\tag 4.6
$$
$$
\{\varphi ,\varphi \}=0=\{\varphi ,\phi \}\tag 4.7
$$

The price one has to pay is that (4.6) is a set of infinite reducible
constraints since
$$
\np \varphi \equiv 0\ \ \ \ \ .\tag 4.8
$$

Over the manifold defined in (4.2) and (4.6) constraints (4.3) become reducible
since
$$
\np \phi = 0 \tag 4.9
$$
holds identically. The manifold defined by (4.2) and (4.3) may thus be
equivalently described by the first class infinite reducible constraint (4.2),
(4.6) and the reducible second class constraints (4.3). In order to write the
effective action of the system one has to include the ghost fields adequate
for the first class reducible constraints $\varphi$ and devise a method
to handle the second class reducible constraints (4.3).

This decomposition of the constraints into first class and reducible second
class ones, over the manifold of first class constraints, is best
understood by introducing tranverse + longitudinal $(T+L)$ decomposition of
the geometrical object. Although $\np$ is not an invertible matrix it serves
to define such decomposition of spinors. Due to the identity
$$
\np \gamma^+ + \gamma^+ \np = P_{-}{\Lambda 1}\tag 4.10a
$$
for any spinors $\xi$ one has the (T+L) decomposition
$$
\xi = \xi_\top +\gamma^+ \xi_L \tag 4.10b
$$
with
$$
\np \xi_\top = 0\ \ \ \ \ .\tag 4.10c
$$
$\xi_L$ is not uniquely defined but $\gamma^+\xi_L$ and $\xi_\top$ are
uniquely determined.

Equation (4.9) imposes that the longitudinal part of $\phi$, more precissely
$\gamma^+\phi_L$, be identically zero over the manifold of first class
constaints. The true content of the reducible constraints (4.3) is then only
$$
\phi_\top = 0\ \ \ \ \ .\tag 4.11
$$

Let us translate the above  situation to a general notation. We have a
constrained system with  Hamiltonian $H_0$ subject to a set of reducible
constraints $\phi_{a_1}$  $(a_1 =1,\cdots ,n)$ and a set of first class
constraints  $\varphi_i$ $(i=1,\cdots ,k)$ which we omit in the explicit
construction that  follows. We limit ourselves to remark on the modifications
to be done when  included. So we have,
$$\align
\phi_{a_1}&=0\tag 4.12\\
a_{a_2}^{a_1}\phi_{a_1}&=0\ \ \ \ \ \ \ a_1=1\cdots n,\ \ \ a_2=1\cdots m
\ \ \ \ \ .\tag 4.12a
\endalign
$$

We will not suppose $a_{a_2}{}^{a_1}$ to be of maximal rank. Instead we will
impose that a (T+L) decomposition similar to (4.10) is allowed.

We have then for any object $V_{a_1}$
$$\align
V_{a_1} &=V^\top_{a_1}+A_{a_1}^{a_2}V_{a_2}^L\tag 4.13a\\
a^{a_1}_{a_2}V_{a_1}^\top =0,&\ \ \ \ \ \ \ V_{a_2}^L=a^{a_1}_{a_2}V_{a_1}
\endalign
$$
and for any object $W^{a_1}$
$$\align
W^{a_1} &=W_\top^{a_1}+a^{a_1}_{a_2}W_L^{a_2}\tag 4.13b\\
A^{a_2}_{a_1}W_\top^{a_1}=0,&\ \ \ \ \ \ \ W_L^{a_2}=A^{a_2}_{a_1}W_{a_1}
\ \ \ \ \ .
\endalign
$$

It follows that
$$
V_{a_2}^L = a^{a_1}_{a_2}A^{b_2}_{a_1}V_{b_2}^{L} ,\ \ \ \ \ \ \
W_L^{a_2}=A^{a_2}_{a_1}a^{a_1}_{b_2}W_L^{b_2}\ \ \ \tag 4.13c
$$
and
$$
W^{a_1}V_{a_1} =W^{a_1}_\top V_{a_1}^\top +W_L^{a_2}V^L_{a_2}
\ \ \ \ \ .\tag 4.13d
$$

In the irreducible case ($a^{a_1}_{a_2}$ of a maximum rank) $A^{a_2}_{a_1}$
is the inverse of $a_{a_2}^{a_1}$. In the finite reducible case this
decomposition may always be done in a unique way for a given pair $A,a$.
For infinite reducible system, we will assume that there exists such a
decomposition.The constraints (4.12) are second class in the sense that they
have an invertible Poisson Bracket matrix in the transverse sub-space.

Following Ref. [7] and [8] let us enlarge the phase space using a set of
auxiliary variables $\xi^{a_1}$ and $\eta_{b_1}$ conjugate to each other.
We also introduce the combinations
$$\align
\Phi_{a_1} &=\eta_{a_1}-{1\over{2}}\omega_{a_1b_1}(p,q)\xi^{b_1}\\
\ov{\Phi}_{a_1} &=\eta_{a_1}+{1\over{2}}\omega_{a_1b_1}(p,q)\xi^{b_1}
\ \ \ \ \ .\tag 4.14
\endalign
$$

Here $\omega_{ab}$ is an antisymmetric matrix with vanishing Poisson Bracket
with itself to be fixed by the procedure. $\ov{\Phi}$ and $\Phi$ satisfy
$$\align
\{\Phi_{a_1},\Phi_{b_1}\}\ &=-\omega_{a_1b_1}\\
\{\ov{\Phi}_{a_1},\ov{\Phi}_{b_1}\}\ &=\omega_{a_1b_1}\\
\{\Phi_{a_1},\ov{\Phi}_{b_1}\}\ &=0\ \ \ \ \ .\tag 4.15
\endalign
$$

In order to introduce only the complications necessary to deal with the case
of the BS superparticle we will suppose in the following that
$\omega_{a_1b_1}$ is transverse
$$
a^{a_1}_{a_2}\omega_{a_1b_1}=0\ \ \ \ \ \ a_1=1\cdots n
\ \ \ \ \ \ a_2=1\cdots m\tag 4.16
$$
and invertible in the transverse space. Now we extend the constraints in
the enlarged space to
$$
\wt{\phi}_{a_1}=\phi_{a_1}+V^{c_1}_{a_1}\Phi_{c_1}=0\tag 4.17
$$
where  $V^{c_1}_{a_1}(q,p)$ is also to be fixed. In general the first class
constraints $\varphi$ may  also have to be extended in order the complete set
of extended constrains be first class. In this case of the superparticle,
however, the extension is not necessary. We assume $V_{a_1}^{b_1}$ to
be invertible. In this case we impose the constraints (4.17) to be irreducible,
first class and with structure functions at most linear in $\Phi_{a_1}$. We
then have
$$
\{\wt{\phi}_{a_1},\wt{\phi}_{b_1}\}=U_{a_1b_1}^{c_1}\wt{\phi}_{c_1}
=-2(u_{a_1b_1}^{c_1}+v_{a_1b_1}^{c_1d_1} \Phi_{d_1})\wt{\phi}_{c_1}\ \ \ \ \
.\tag 4.18
$$

The structure functions $U_{a_1b_1}^{c_1}$ may depend on the phase space
variables $p$ and $q$. Substitution of (4.17) in (4.18) yields.
$$\align
\{\phi_{a_1},\phi_{b_1}\} &-V_{a_1}^{c_1}V_{b_1}^{d_1}\omega_{c_1d_1}+
2u_{a_1b_1}^{c_1}\phi_{c_1}=0\\
\{\phi_{a_1},V_{b_1}^{c_1}\}+\{V_{a_1}^{c_1},\phi_{b_1},\}
&+2v_{a_1b_1}^{d_1c_1}\phi_{d_1}+ 2u_{a_1b_1}^{d_1}V_{d_1}^{c_1}=0\\
\{V_{a_1}^{c_1},V_{b_1}^{d_1}\}+\{V_{a_1}^{d_1},V_{b_1}^{c_1}\}
&+2V_{e_1}^{c_1}v_{a_1b_1}^{e_1d_1}+ 2V_{e_1}^{d_1}v_{a_1b_1}^{e_1c_1}=0
\ \ \ \ \ .\tag 4.19
\endalign
$$

We suppose here that
$$
\{\phi_{a_1},\Phi_{1a_1}\}=0,\ \ \ \{V_{1a_1}^{b_1},\Phi_{1c_1}\}=0
\ \ \ \ \ . \tag 4.20
$$

Let us suppose that we are able to find a solution to (4.20) with all the
required conditions. In order to demonstrate the equivalence of our system
in the enlarged phase  space to the original system we have to impose
additional restriction besides  (4.17). A counting of the degrees of
freedom suggests which ones should be  chosen. The original model has $2N$
phase  space variables $p,q$ restricted by ($n-m_L$) transverse constraints
with  $m_L$ the rank of $a^{a_2}_{a_1}$. The enlarged model has $2N$
variables  $p,q$ and $2n$ variables $\xi$, $\eta$ restricted by $n$
constraints  $\wt{\phi}_{a_1}$ and $n$ gauge fixing conditions
$\wt{\chi}_{a_1}$. To match  we need ($n-m_L$) additional constraints. We
take them to be $$ \ov{\Phi}_{a_1}^\top =0\ \ \ \ \ .\tag 4.21
$$
Since
$$
[\ov{\Phi}_{a_1}^\top ,\ov{\Phi}_{a_2}^\top ]=\omega_{a_1a_2}^\top \tag 4.22
$$
the constraints (4.21) are in our hypothesis second class.The advantage of
this formulation is that he field dependence in $\omega^{\top}_{a_1a_2}$
(whose determinant will appear in functional measure) may be simpler than in
$\{\phi_{a_1},\phi_{b_1}\}$ since  $V_{a_1}^{a_2}$ may be also a field
dependent object. This justifies the enlarging of the phase space and the
modification of the constraints. Moreover iterating the process one can hope
to obtain a field independent functional measure an a pure gauge model. For
the BS superparticle, as we will show below infinitely many iterations are
needed to this end, but in other cases only finite steps may be necessary.

A gauge invariant extension of the hamiltonian $H_0$ may be written in the
form [11]
$$
\wt{H}=H_0+h^{a_1}\Phi_{a_1}\ \ \ \ \ .\tag 4.23
$$
$h^{a_1}$ is fixed imposing
$$
\{\wt{H},\wt{\phi}_{a_1}\} =W_{a_1}^{b_1}\wt{\phi}_{b_1}\ \ \ \ \ .\tag 4.24
$$

Introducing the ghost variables $C^{a_1}$ and $\mu_{a_1}$ the BRST operator
is obtained by solving [6]
$$\align
\{\Omega ,\Omega \}&=0\tag 4.25a\\
\left. {\partial\Omega\over{\partial C^{a_1}}}\right |_{\mu =0} & =
\wt{\phi}_{a_1}\ \ \ \ \ .\tag 4.25b
\endalign
$$

It takes the form
$$
\Omega =C^{a_1}\wt{\phi}_a +C^{a_1}U_{a_1b_1}^{c_1}C^{b_1}\mu_{c_1}+\cdots
\tag 4.25c
$$
with a non-trivial tail, when the algebra of first class constraints has
structure functions of a higher order. In the general case when first class
constraints $\varphi$ are also present, one has to include, of course,
associated ghost fields and condition (4.25b) must also be satisfied for the
extended first class constraints $\wt{\varphi}$.

The extended hamiltonian is then obtained by solving [5]
$$\align
\{ \wh{H},\Omega \}&=0\\
\left. \wh{H}\right |_{\mu =0} &=\wt{H}\ \ \ \ \ .\tag 4.26
\endalign
$$

The BRST invariant effective action in a phase space representation is given
by [5]
$$
S_{eff}=<p\dot{q}+\mu_{a_1}\dot{C}^{a_1}+\eta_{a_1}\dot{\xi}^{a_1}-
\wh{H}+\wh{\delta}(\lambda^{a_1}\mu_{a_1})+\wh{\delta}(\ov{C}_{a_1}
\chi^{a_1})>\tag 4.27
$$
where $\chi^{a_1}$ are the gauge fixing conditions and $\wh{\delta}$ is
defined by
$$
\wh{\delta}F=[\Omega ,F]\tag 4.28
$$
for any function $F$ of the canonical variables of the enlarged
superphase-space. For the non-canonical sector we have
$$\align
\wh{\delta}\lambda^{a_1} &=\theta^{a_1}\ \ \ ,\ \wh{\delta}\theta^{a_1}=0
\tag 4.29a\\
\wh{\delta}\ov{C}_{a_1} &=B_{a_1}\ \ \ ,\ \wh{\delta}B_{a_1}=0\tag 4.29b\\
\delta\mu_{a_1}&=\wt{\phi}_{a_1}\ \ \ \ \ .\tag 4.29c
\endalign
$$

We claim that the gauge invariant system defined by (4.27) and constrained by
(4.21) is canonically equivalent to the original system. To prove this, we will
show that with an adequate gauge fixing condition one can reduce the path
integral corresponding to the enlarged system to the Senjanovic-Fradkin
expression for the original system.

The classical gauge transformation law for $\xi$ is
$$
\delta\xi^{a_1}=\{\xi^{a_1},\epsilon^{b_1}\wt{\phi}_{b_1}\}=V_{b_1}^{a_1}
\epsilon^{b_1}\tag 4.30
$$
where $\epsilon^{b_1}$ are the infinitesimal parameters of the transformation.

We may then choose the gauge conditions
$$
\chi^{a_1}=\xi^{a_1}\ \ \ \ \ .\tag 4.31
$$

Using (4.25), (4.28) and (4.29) we have
$$\align
\wh{\delta}(\ov{C}_{a_1}\chi^{a_1}) &=B_{a_1}\chi^{a_1}-\ov{C}_{a_1}
{\delta\chi^{a_1}\over{\delta\epsilon^{b_1}}}C^{b_1}+O(\mu )\\
&=B_{a_1}\xi^{a_1}-\ov{C}_{a_1}V^{a_1}_{b_1}C^{b_1}+O(\mu )\tag 4.32
\endalign
$$
where $O(\mu )$ may appear if the structure functions depend explicitly on
the phase space coordinates. We also have
$$\align
\wh{\delta}(\lambda^{a_1}\mu_{a_1}) &=\lambda^{a_1}\wt{\phi}_{a_1}+
\theta^{a_1}\mu_{a_1}\\
&=\lambda_\top^{a_1}\wt{\phi}_{a_1}+
\lambda_L^{a_2}a_{a_2}^{a_1}\wt{\phi}_{a_1}+\theta^{a_1}\mu_{a_1}\\
&=\lambda_\top^{a_1}\wt{\phi}^\top_{a_1}+
\lambda_L^{a_2}a_{a_2}^{a_1}V_{a_1}^{b_1}\Phi_{b_1}+\theta^{a_1}\mu_{a_1}
\ \ \ \ \ .\tag 4.33
\endalign
$$

The functional integral is
$$
(\chi )=\int \D z\delta (\ov{\Phi}^\top )(det \omega^\top )^{1/2}e^{-Seff}
\tag 4.34
$$
where $\D z$ is the Liouville measure
$$
\D z=\D p\D q \D C \D \ov{C}\D \mu \D B\D \theta \D \eta \D \xi \tag 4.35
$$

Integrating in $\theta$ one gets $\delta (\mu )$ so that in particular
$O(\mu )$ in (4.32) does not contribute. Integrating in $B_{a_1}$,
$\ov{C}_{a_1}$,and $\lambda^{a_2}_L$ and using Eq.(4.19) the factor in the
measure of (4.34) becomes
$$\align
&(det \omega^\top )^{1/2}\delta (\ov{\Phi}^\top )\delta (\xi )\delta
(V_{\top b_1}^{\top a_1}C_\top^{b_1})\delta (a_{b_2}^{b_1}V_{b_1}^{a_1}
A_{a_1}^{a_2}C_L^{b_2})\delta (a_{a_2}^{a_1}V_{a_1}^{b_1}\Phi_{b_1})=\\
&(det \omega^\top )^{1/2}\delta (\eta^\top )\delta (\xi )\delta
(V_{\top b_1}^{\top a_1}C_\top^{b_1})\delta (a_{b_2}^{b_1}V_{b_1}^{a_1}
A_{a_1}^{a_2}C_L^{b_2})\delta (a_{a_2}^{a_1}V_{a_1}^{b_1}A_{b_1}^{a_2}
\eta_{a_2}^L )\tag 4.36
\endalign
$$

In (4.36) the arguments of the last two factors have opposite statistics.
Hence
$$
\delta (a_{b_2}^{b_1}V_{b_1}^{a_1}A_{a_1}^{a_2}C_L^{b_2})
\delta (a_{a_2}^{a_1}V_{a_1}^{b_1}A_{b_1}^{a_2}\eta_{b_2}^L)=
\delta (C_L^{b_2})\delta (\eta^L_{a_2})\ \ \ \ \ .\tag 4.37
$$

This can be taken as valid even in the case of $a^{a_1}_{a_2}$ and
$A^{b_1}_{b_2}$ being non invertible. The factor in the measure reduces to
$$
(det \omega^\top )^{1/2}\delta (\eta )\delta (\xi )\delta (C)det
V_{\top}^\top \ \ \ \ \ .\tag 4.38
$$

Now we note from (4.19) that
$$
(det \omega^\top )^{1/2}det V_{\top}^\top =(det \{\phi_\top ,\phi_\top \} )
\ \ \ \ \ .\tag 4.39
$$

Doing the trivial integrals in $\eta$, $\xi$ and $C$, we finally obtain
$$
I=\int \D q \D p \D \lambda^\top (det \{\phi_\top ,\phi_\top \})exp-
<p\dot{q}-H+\lambda_\top \phi^\top >\tag 4.40
$$
which is the correct Senjanovic-Fradkin expression of the functional integral
of this system.

The discussion above only supposes the uniqueness of decomposition (4.13).
For the superparticle we identify $a^{a_1}_{a_2}$ with $\not{P}$ and
$A^{a_2}_{a_1}$ with $\gamma^+/P_{-}$. The $T+L$ decomposition is given by
(4.11). Solving (4.19) by taking $V$ as a Dirac $\delta$ we have
$$
\omega=-2\np \ \ \ \ \ .\tag 4.41
$$

In terms of the auxiliary Majorana spinors $\eta_1$ and $\xi_1$ we have
$$\align
\Phi_1 &=\eta_1+\np \xi_1 \tag 4.42a\\
\ov{\Phi}_1 &=\eta_1-\np \xi_1 \ \ \ \ \ .\tag 4.42b
\endalign
$$

The enlarged constraints (4.17) are in this case
$$
\wt{\phi}_0=\eta -\np \xi +\Phi_1\ \ \ \ \ .\tag 4.43
$$

The additional restrictions corresponding to (4.21) are
$$
\ov{\Phi}_1^\top =0\ \ \ \ \ .\tag 4.44
$$
Since we choose $V_{a_1}^{b_1}$ to be field independent the factor
$det\{\ov{\Phi}^\top_1,\ov{\Phi}^\top_1\}^{1/2}$ in the measure of functional
integral appears in principle in this case as problematic as the factor
$det\{\phi^\top ,\phi^\top \}^{1/2}$ in the  direct approach. Nevertheless we
observe that constraint (4.44) is equivalent to reducible constraint
$$
\align
\np \ov{\Phi}_1|_{\text{first class}} &=\np \eta_1 \equiv 0 ,\tag 4.45a\\
\ov{\Phi}_1 &=0 . \tag 4.45b
\endalign
$$

We iterate now the process and introduce $\xi_2$, $\eta_2$ and $\omega_2$.
We obtain again
$$\align
\omega_2 &=-2\np \\
\Phi_2 &=\eta_2+\np \xi_2 \\
\ov{\Phi}_2 &=\eta_2-\np \xi_2 \tag 4.46
\endalign
$$
and we have the new constraints
$$\align
\wt{\phi}_1 &=\ov{\Phi}_1+ \Phi_2 \tag 4.47a\\
\ov{\Phi}^\top_2 &=0 \tag 4.47b
\endalign
$$

For the same reason as above we take instead of (4.47b) the reducible
constraint
$$\align
\np \ov{\Phi}_2|_{\text{first class}} &=\np \eta_2 \equiv 0 \tag 48a\\
\ov{\Phi}_2 &=0 .\tag 4.48b
\endalign
$$
and continue the process. After $\ell$ steps we have
$$\align
\wt{\phi}_{i-1} &=\ov{\Phi}_{i-1}+\ov{\Phi}_i\ \ \ \ i=1,\cdots ,\ell\\
\ov{\Phi}^\top_{\ell} &=0 \tag 4.49
\endalign
$$
with $\ov{\Phi}_0 \equiv \phi$.

At this level the classical action may be written in terms of the canonical
variables in the form
$$
S_\ell =<P_\mu \dot{x}^\mu +\sum_{i=0}^\ell \eta_i \dot{\xi}^i+\lambda
P^2+\sum_{i=0}^{\ell -1}\ov{\psi}^i\np \eta_i +\sum^{\ell}_{i=0}\lambda^i
\wt{\phi}_{i-1}>.\tag 4.50a
$$
subject to the second class constraints
$$
\ov{\Phi}^\top_\ell =0.\tag 4.50b
$$
In (4.50a) $\ov{\psi}^i$ is the Lagrange multiplier associated to the i-esim
analog to (4.45a) and (4.48a). The constraints $\wt{\phi}_\ell$ in (4.49) are
irreducible, the other constraints being infinite reducible. At each level
$\ell$ the formulation (4.50) is not Lorentz covariant due to the transverse
projection in (4.50b).

In order to avoid this problem we may introduce infinite auxiliary fields. We
then obtain
$$
S_\infty =<P_\mu \dot{x}^\mu +\sum_{i=0}^\infty \eta_i \dot{\xi}^i +
\lambda P^2+\sum_{i=0}^{\infty}\ov{\psi}^i\np \eta_i +
\sum^{\infty}_{i=0}\lambda^i\wt{\phi}_{i-1}>.\tag 4.51
$$

This is the action proposed by Kallosh in [12] which is associated to the
BRST charge with the correct cohomology for the BSSP. The effective action
associated to (4.51) may be truncated at any level $\ell$ by imposing the gauge
fixing conditions (4.31) for $i=\ell +1,\cdots ,\infty$, and the effective
action associated to (4.50) is regained. In the limit case $\np \eta_i=0$ and
$\wt{\phi}_i=0$ $i=0,\cdots ,\infty$ are regular infinite reducible first
class constraints. Other approaches for the quantization of the
superparticle may be found in [14]
\vskip 5mm
\noi
{\bf 5. On the Covariant Quantization of GSSS}

\vskip 3mm

We are now going to extended the approach of section 4  in order to
apply it to the GSSS. The constraints (2.5) can be decoupled into left and
right sectors $$\align
\phi_- &\equiv \phi_1-\phi_2 \tag 5.1a\\
F^1 &=0\tag 5.1b
\endalign
$$
and
$$\align
\phi_+ &\equiv \phi_1+\phi_2 \tag 5.2a\\
F^2 &=0\tag 5.2b
\endalign
$$

We analyse here the problem related to sector (5.1), that is the extension
of the phase of section 2 in order to obtain a regular formulation with first
class constraints only which reduces off-shell to the left sector of GSSS,
with the right canonical functional measure.

We consider the reducible set of constraints
$$\align
H_- &= \phi_-+2F^1\theta^{1'}=0 \tag 5.3a\\
F^1 \Gamma_1&=0\tag 5.3b\\
F^1 &=0\tag 5.3c
\endalign
$$
The $\Gamma_1$ martrices satisfy the property
$$
\Gamma_1\Gamma_1=2H_-{\text{\bf{\catorce 1}}},\tag 5.4
$$
hence the first class constraints (5.3b) are infinite reducible with respect
to $\Gamma_1$. The procedure of section 5 may be generalized as follows, we
consider the set of restrictions
$$\align
H_- &=0 \tag 5.5a\\
\chi^- &=0 \tag 5.5b\\
F^1\Gamma_1 &=0 \tag 5.5c\\
\chi_1 &=0 \tag 5.5d\\
F^1 &=0 \tag 5.5e
\endalign
$$
where $\chi^-$ and $\chi_1$ are covariant gauge fixing functions associated
to $\phi_-$ and $F^1\Gamma_1$ respectively.

(5.5a), (5.5d) are a reducible subset which we shall denote $\varphi_i$,
while $\chi^-$, $\chi_1$ are going to be denoted by $\chi^j$. The
constraints (5.5) are  seconds class . We now extended (5.5) to obtain a set
of first  class ones; we enlarge the phase space with auxiliary canonical
coordinates  $(\mu_j,\rho^j)$ and $(\eta ,\xi )$ as follows
$$\align
\wt{\varphi}_i &=\varphi +M_i{}^j\mu_j+N_{ij}\rho^j \tag 5.6a\\
\wt{\chi}^j &=\chi^j+L^{ji}\mu_i+\rho^j \tag 5.6b\\
\wt{F}^1 &=F^1+\Phi +P^i\mu_i+Q_j \rho^j \tag 5.6c
\endalign
$$
where $\Phi =\eta +\omega \xi$ (see section 4).

It can be shown that associated to system (5.1) or (5.2) there exist
$M$, $N$, $P$, $Q$ such that (5.6) are first class constraints. Moreover,
(5.6a) is reducible with respect to $\Gamma_1$.

It can be also proven that by imposing the partial gauge fixing conditions
$\mu_i=0$ and $\chi^j=0$ associated to (5.6a) and (5.6b) respectively, the
functional measure corresponding to (5.6a) and (5.6b) reduces exactly to the
Fradkin-Senjanovic measure associated to the constraints $\phi_-$ and
$F^1\Gamma_1$.

Constraints (5.6) may be reexpressed as follows
$$\align
\wh{\wt{\varphi}}_i &=\wt{\varphi}_+\wt{M}_i{}^j\mu_j \\
\wh{\wt{\chi}}^j &=\wh{\chi}^j+\rho^j \\
\wh{\wt{F}}^1 &=\wt{F}^1+\Phi \tag 5.7
\endalign
$$
where $\wh{\varphi}_i$, $\wh{\chi}^j$ and $\wt{F}^1$ are independent of the
auxiliary fields $\rho$, $\mu$ and $\Phi$.

We consider now the set of first class constraints (5.7) together with the
second class constraint
$$
\ov{\Phi}+R^j\mu_j+S_j\rho^j=0\tag 5.8
$$
where $\ov{\Phi}=\eta -\omega\xi$, and $R^j$, $S_j$ can be determined in
order that (5.7) are first class constraints. This equation may be
reexpressed, by taking linear combinations of (5.7) and (5.8), as
$$
\wh{\ov{\Phi}}=0\tag 5.9
$$
with no dependence on the auxiliary fields.

Eq. (5.7) and (5.9) are the generalization to the superstring case of the
superparticle constraints (4.2), (4.17) and (4.21).

The procedure is now exactly as in section 4. We obtain
$$\align
\wh{\wt{\varphi}}_i &=0\\
\wh{\wt{\chi}}^j &=0\\
\wh{F}^1+\Phi &=0\\
\wh{\Phi}+\Phi_1 &=0\\
\wh{\Phi}_1+\Phi_2 &=0\\
\vdots \ \ \ & \\
\wh{\Phi}_n+\Phi_{n+1}=0\ \ \ \ \ n\to \infty \tag 5.10
\endalign
$$
which are all first class.

The reduction procedure which we have explicitly shown in section 4 for the
superparticle problem can be extended to this case. Moreover it can be
shown  that it is independent of the $\chi^j$ function used in (5.10).

It is interesting to compare the formulation of this section with  the
one in section 4 where no use was made of this constraints
$\wt{\chi}=0$. In those cases the original first class sector commutes with
everything else, in particular with $\omega$ and hence with any extension of
the original second class sector. There is no need of extending the original
first class sector. However if one introduces the $\wt{\chi}=0$ restriction
as for the more general GSSS case, it can be shown that the $(\rho ,\mu)$
auxiliary sector decouples and can be eliminated by functional integration
ending up with the original  formulation of section 4.

It remains to analyse the admissible set of gauge fixing conditions
associated to (5.10) as well as the explicit construction of the BRST
charge  which nevertheless will have necessarily the correct cohomology
since if is going to be obtained  from a regular canonical formulation which
reduces correctly to the GSSS. This will be presented with a detailed
analysis of our formulation elsewhere [15]
\vskip 5mm

\item{}{\bf References}
\vskip .3cm

\item{[1]}M. Green and J. Schwarz, Phys. Lett. {\bf 136B} (1984) 367.
\item{[2]}M. B. Green and J. H. Schwarz, Nucl. Phys. {\bf 243} (1984) 475;
\item{}L. Brink, in: {\bf Unified String Theories}, eds. M. Green and D.
Gross (World Scientific, Singapore, 1986).
\item{[3]}J. Greensite and F. Klinkhammer, Nucl. Phys. {\bf B281} (1987) 269.
\item{}A. Restuccia and J. G. Tylor, Int. J. of M. Phys. {\bf A8} (1993) 753.
\item{[4]}A. Restuccia and J. G. Tylor, King's College Preprint (1993)
\item{[5]}I. Batalin and E. Fradkin, Phys. Lett. {\bf B122} (1983) 157, ibid
{\bf B128} (1983) 307.
\item{}M. Caicedo and A. Restuccia, Class. Q. Grav. {\bf 10} (1993) 833.
\item{[6]}M. Caicedo and A. Restuccia, preprint (USB) 1993.
\item{[7]}A. Restuccia and J. Stephany,Phys. Lett. {\bf 305B} (1993) 348;Phys.
Rev. {\bf D47} (1993) 3437
\item{}R. Gianvittorio, A. Restuccia and J.Stephany, Mod.Phys. Lett. {\bf
A6}  (1991) 2121.
\item{[8]}I.Batalin and E.S.Fradkin , Nucl. Phys. {\bf 279} (1987) 514.
\item{[9]}V.Rivelles and J.G.Taylor, Phys.Lett. {\bf B104} (1981) 131;
\item{}W.Siegel and M.Rocek, Phys Lett. {\bf B105} (1981) 275;
\item{[10]}T. Hori and K. Kamimura, INS-Rep-474 (1983).
\item{[11]}M.Henneaux and C.Teitelboim in {\it Quantum Mechanics of
Fundamental Systems 2} eds. C.Teitelboim and J.Zanelli, Plenum Press (New
York 1989) p.83
\item{}M. Lledo and A.Restuccia, Preprint USB, 1993; .
\item{[12]}R. Kallosh, Phys. Lett. {\bf B251} (1990) 134.
\item{[13]}I. Batalin, R. Kallosh and A. Van Proeyen in Quantum Gravity
eds. M. Markov, V. Berezin and F. Frolov, World Scientific Singapore.
\item{}M. B. Green and C. M. Hull, Mod. Phys Lett. {\bf A5} (1990) 1399
\item{[14]}Y Eisenberg and S.Solomon, Nuc Phys  {\bf B309} (1988) 709.
\item{}N.Berkovits, Nuc Phys  {\bf B350} (1991) 193
\item{}Y.Eisenberg, Preprint IASSNS-HEP-91/48, Princeton
\item{}N.Ilieva and L.Litov,Preprint IC/92/180 ,Trieste
\item{[15]}A. Restuccia and J. Stephany in preparation.

\bye